\documentclass{iau} 
\usepackage{graphicx}

\title[Gamma-ray bursts: A brief survey of the diversity]
{Gamma-ray bursts: A brief survey of the diversity}

\author[Attila M\'esz\'aros \& Jakub \v{R}\'{\i}pa]
{Attila M\'esz\'aros$^1$ \& Jakub \v{R}\'{\i}pa$^{1,2,3}$}

\affiliation{$^1$Astronomical Institute, Faculty of Mathematics and Physics, 
Charles University, CZ-180 00 Prague 8, V Hole\v{s}ovi\v{c}k\'ach 2, Czech Republic\\ email:{\tt meszaros@cesnet.cz}\\[\affilskip]
$^2$MTA-E\"otv\"os L\'or\'and University, Lend\"ulet Hot Universe Research Group, P\'azm\'any P\'eter s\'et\'any 1/A, Budapest, 1117, Hungary
\\$^3$E\"otv\"os L\'or\'and University, Institute of Physics, P\'azm\'any P\'eter s\'et\'any 1/A, Budapest, 1117, Hungary\\email:{\tt jripa@caesar.elte.hu}}

\pubyear{2018}
\volume{346}
\jname{High-mass X-ray binaries: illuminating the passage from
massive binaries to merging compact objects}
\editors{A.C. Editor, B.D. Editor \& C.E. Editor, eds.TO BE PRECISED}

\begin{document}

\maketitle

\begin{abstract}
The separation of the gamma-ray bursts (GRBs) into short/hard and long/soft subclasses, respectively, 
is well supported both theoretically and observationally.
The long ones are coupled to supernovae type Ib/Ic - the short ones are connected to the merging
 of two neutron stars, where one or even both neutron stars can be substituted by black holes. 
These short GRBs - as merging binaries - can also serve as the sources of gravitation waves,
and are observable as the recently detected macronovae. Since 1998 there are several statistical studies 
suggesting the existence of more than two subgroups. There can be a subgroup
having an intermediate durations; there can be a subgroup with ultra-long durations; 
the long/soft subgroup itself can be divided into two subclasses with respect to the luminosity of GRBs.
The authors with other collaborators provided several statistical studies
in this topic. This field of the GRB-diversity is briefly surveyed in this contribution.
\keywords{cosmology: miscellaneous, cosmology: observations, 
gamma rays: bursts}
\end{abstract}

\firstsection

\section{Introduction}

After the discovery of the gamma-ray bursts (GRBs) (\cite[Klebesadel et al. 1973]{Kle73})
it was shown from the statistical analyses of observational data that they were two types of GRBs 
(\cite[Mazets et al. 1981]{Maz81}). Since that time
the se\-paration of GRBs into subgroups is a blistering problem. 
In this contribution a brief survey of this topic is provided.

\section{Separation into two subgroups}

Since 1981 the separation of GRBs into short/hard and long/soft subgroups was confirmed by several other studies.
For the GRBs, not having directly measured redshifts, the so called hardness can be used in statistical 
tests together with the duration (for a survey and the relevant references see, e.g., \cite[M\'esz\'aros 2006]{Mesz06}).
 
Today it is clear that the short/hard bursts
are given by the neutron star - neutron star mergers, where one or even two neutron stars can be substituted 
by black holes, and are observable as macronovae (\cite[Tanvir et al. 2013]{Tan13}). 
The cosmologically near short/hard bursts can also serve as the source for the detectable gravitational waves 
(\cite[Abbott et al. 2017]{Abb17}). This means that the short/hard GRBs arose at the final stage of compact binaries.

The long/soft bursts are associated with Ib/c type supernovae
(\cite[Hjorth et al. 2003]{Hjo03}). The idea that the GRBs were coupled to supernovae was formulated in essence
simultaneously with the discovery of bursts (\cite[Colgate 1968]{Col68}, \cite[Colgate 1974]{Col74}). 

The separation should be done with respect to the duration at $\simeq 2$ second. But this limiting duration - as a strict one - 
is in doubt, because much longer GRBs were also observed, which resemble the short/hard GRBs (\cite[Gehrels et al. 2006]{Ge06}).
This means that it is better to speak about two basic types of GRBs (\cite[Kann et al. 2010]{Ka10}).

\section{Three subgroups?}

In year 1998 two simultaneous papers declared the existence of a third subgroup (\cite[Mukherjee et al. 1998]{Mukh98}, \cite[Horv\'ath 1998]{Ho98}).
This claim came from the statistical studies of the dataset of BATSE instrument being on the Compton Gamma Ray Burst 
Observatory\footnote{https://heasarc.gsfc.nasa.gov/docs/cgro/index.html}. Since that time several other papers declared the same result for
the BATSE dataset (\cite[Horv\'ath 2002]{Ho02}, \cite[Hakkila et al. 2003]{Ha03}, \cite[Hakkila et al. 2004]{Ha04},
\cite[Horv\'ath et al. 2006]{Ho06}). This third subgroup should have an intermediate duration. It was found also at the
Swift dataset\footnote{https://swift.gsfc.nasa.gov} (\cite[Veres et al. 2010]{Ve10}). For the 
RHESSI\footnote{https://hesperia.gsfc.nasa.gov/rhessi3} satellite the existence of third 
subgroup was also declared (\cite[\v{R}\'{\i}pa et al. 2012]{Ri12}). On the other hand, no intermediate subgroup 
was found in the Fermi's\footnote{https://fermi.gsfc.nasa.gov} observations (\cite[Tarnopolski 2015]{Ta15}, \cite[Narayana Bhat et al. 2016]{Bath16}). 
Similarly, no third subgroup is declared to exist both in the Suzaku\footnote{http://global.jaxa.jp/projects/sat/astro\_e2/index.html}
database (\cite[Ohmori et al. 2016]{Oh16}), and in the 
Konus/WIND\footnote{http://www.ioffe.ru/LEA/kw/} catalog (\cite[Tsvetkova et al. 2017]{Tsv17}).

It must be added that even in the case, when the three subgroups are found by statistical tests, it is not sure that there are
really three astrophysically different phenomena. Different biases, selection effects, etc...
can play a role (\cite[Hakkila et al. 2003]{Ha03}, \cite[Tarnopolski 2016]{Ta16}).
For example, in the Swift database the third group is found by tests, but
a more detailed study shows that the third group is given by the so-called X-Ray Flashes (XRFs) - which are in essence 
long GRBs (\cite[Veres et al. 2010]{Ve10}). But, on the other hand, in some cases it is claimed that the third subgroup
cannot entirely be given by long GRBs. For example, for the BATSE and mainly for the RHESSI database
the identification of the intermediate GRBs with XRFs cannot be done (\cite[\v{R}\'{\i}pa \& M\'esz\'aros 2016]{Ri16}).
  
\section{Four or even more subgroups?}

There are studies claiming the existence of other subgroups - being not identical - to the intermediate one.

For example, in the BATSE database there were hints for the separation of the long GRBs themselves into the harder and softer parts
(\cite[Pendleton et al. 1997]{Pe97}). 

The longest GRBs can also form an extra - ultra-long - subgroup (\cite[Tikhomirova \& Stern 2005]{Ti05},
\cite[Virgili et al. 2013]{Vir13}, \cite[Levan et al. 2014]{Le14}). Because there are only few cases in this subgroup, from the statistical 
point of view this subgroup hardly can be declared as an astrophysically different phenomenon.

Recently, in the Fermi database five subgroups were found (\cite[Acuner \& Ryde 2018]{Ac18}). The long subgroup should be further separated.
Theoretically, it is thought that seven different subgroups should exist (\cite[Ruffini et al. 2018]{Ru18}).

\section{Newest studies on the samples with known redshifts}

Because the number of GRBs, which have directly measured redshifts from the afterglows observations, is increasing at the last years
it is already possible to study the diversity of GRBs also from other intrinsic quantities. 
The intrinsic luminosity ($L_{iso}$) and the intrinsic total emitted energy ($E_{iso}$) can be calculated for a given GRB, if its redshift is known.
Then these quantities can also be studied instead of hardness.

If such a testing is provided (\cite[Levan et al. 2014]{Le14}) on the duration vs. $L_{iso}$ ($E_{iso}$, respectively) plane, one should obtain 
the subgroups, too. On Fig.2 of Levan et al. (2014) such a study is done. Several possible subgroups are seen beyond the long and short
GRBs (soft gamma repeaters, low luminosity long GRBs, ultra-long GRBs, tidal disruption events). On the other hand, there is no intermediate
subgroup.

\begin{figure}[h]
\begin{center}
\includegraphics[width=0.49\textwidth]{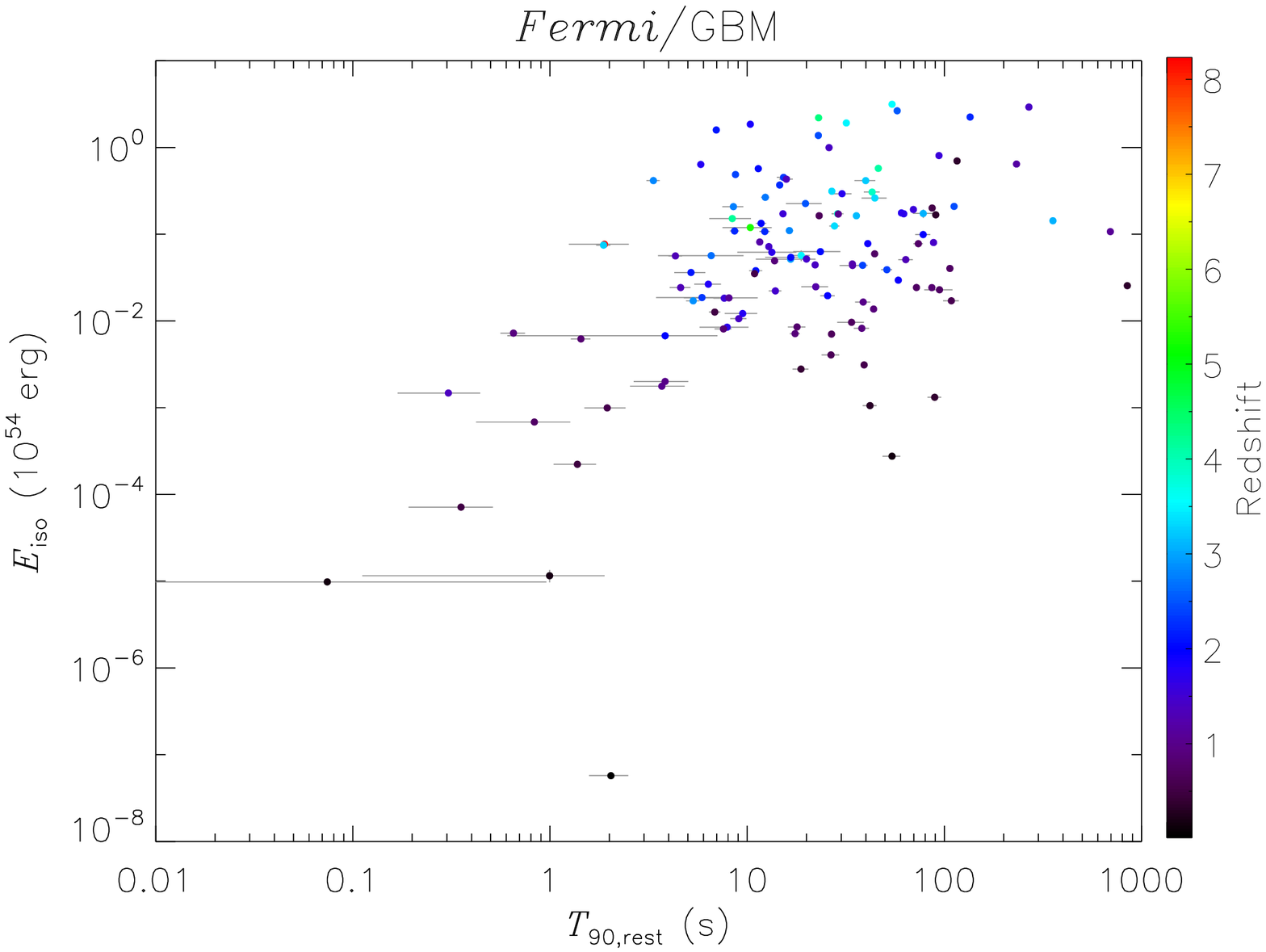}
\includegraphics[width=0.49\textwidth]{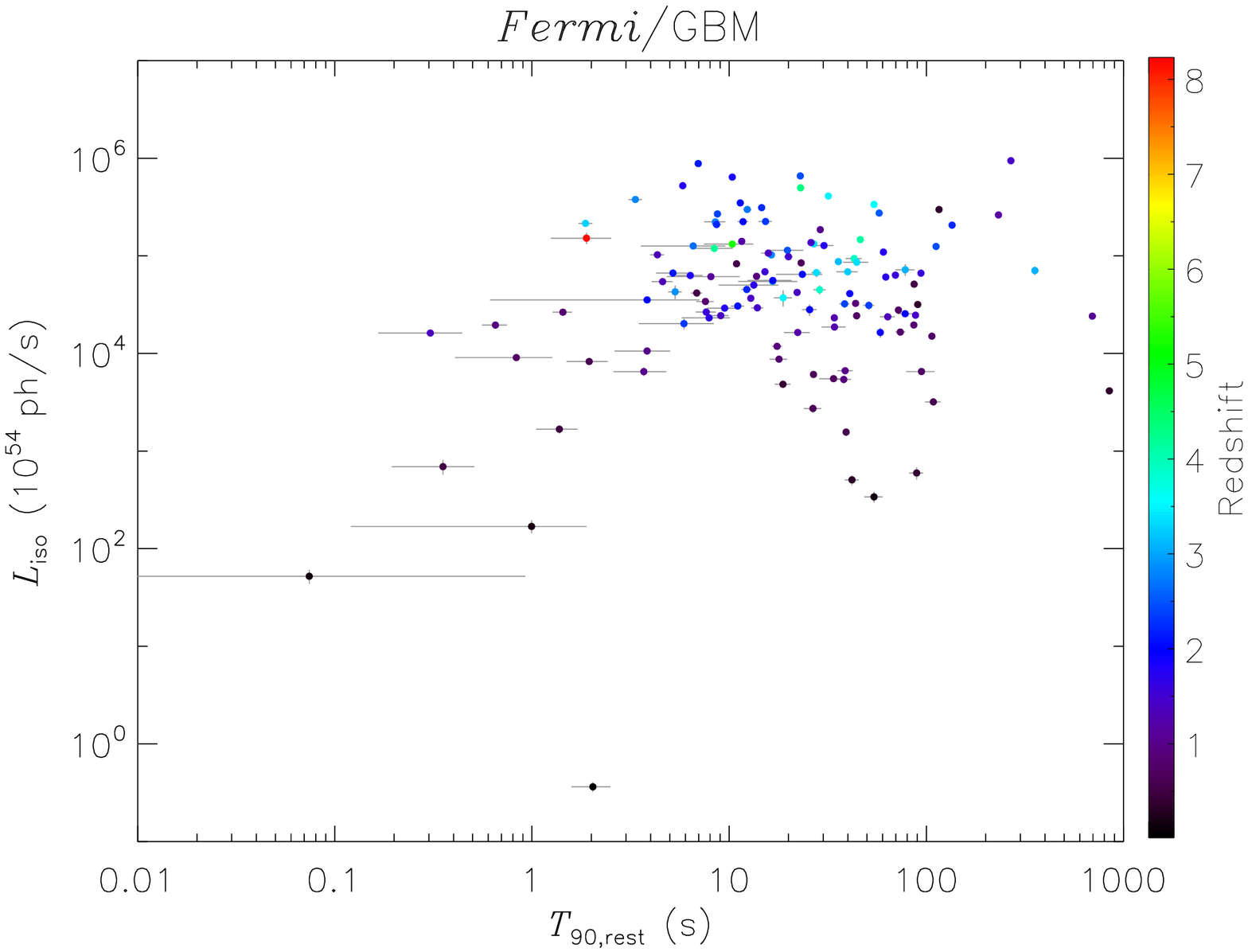}
 \caption{Plotted are 124 {\em Fermi}/GBM GRBs with known redshifts. Isotropic equivalent energy $E_{iso}$ (calculated from the fluence in the 10-1000\,keV band) and isotropic equivalent luminosity $L_{iso}$ (calculated from the 1024 ms peak photon fluxes in the 10-1000\,keV band), respectively, versus duration $T_{90,rest}$ in the rest frame of a GRB are shown.}
   \label{fig1}
\end{center}
\end{figure}

\begin{figure}[h]
\begin{center}
\includegraphics[width=0.49\textwidth]{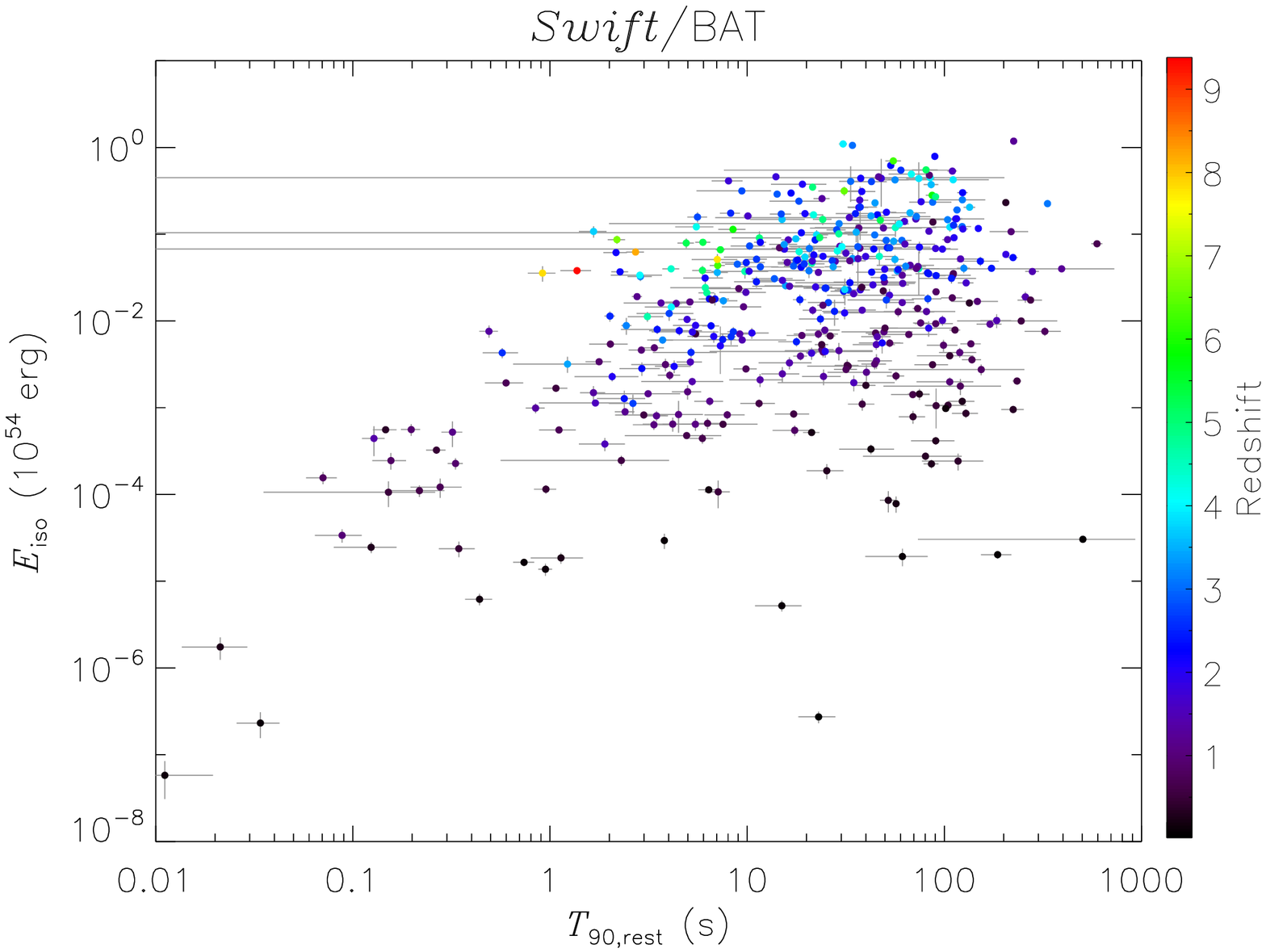}
\includegraphics[width=0.49\textwidth]{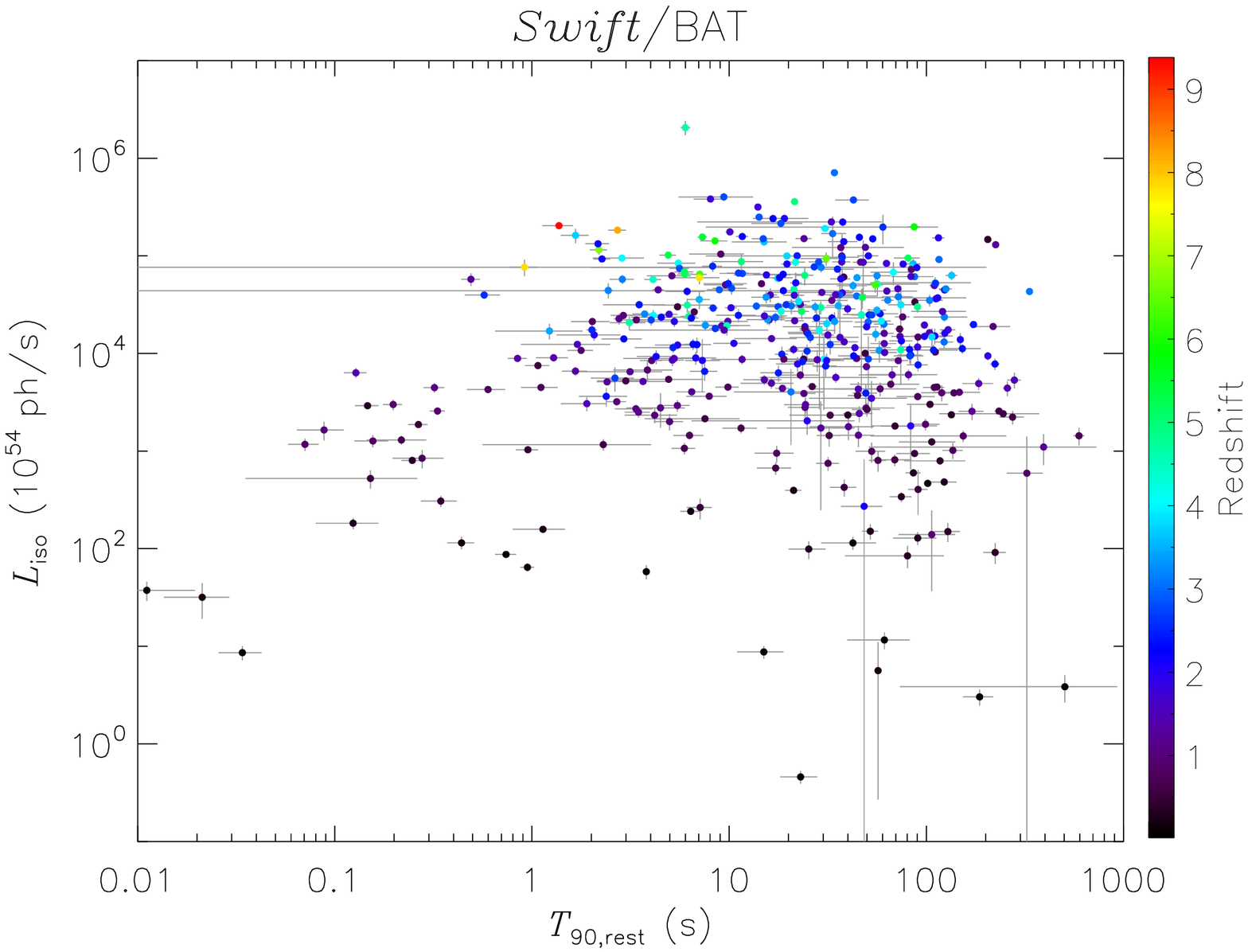}
 \caption{Plotted are 412 {\em Swift}/BAT GRBs with known redshifts. Isotropic equivalent energy $E_{iso}$ (calculated from the fluence in the 15-350\,keV band) and isotropic equivalent luminosity $L_{iso}$ (calculated from the 1 s peak photon fluxes in the 15-350\,keV band), respectively, versus duration $T_{90,rest}$ in the rest frame of a GRB are shown.}
   \label{fig2}
\end{center}
\end{figure}

We present here the preliminary results of a similar effort. We examined data samples of GRBs with measured redshifts from {\em Fermi}/GBM and {\em Swift}/BAT instruments.
The {\em Fermi}/GBM sample was created using the FERMIGBRST
catalog\footnote{https://heasarc.gsfc.nasa.gov/W3Browse/fermi/fermigbrst.html}.
It contains 124 GRBs with the first burst GRB 080804 and the last one GRB 180314A (see Fig.~\ref{fig1}). The {\em Swift}/BAT sample was
created using the {\em Swift} BAT Gamma-Ray Burst Catalog\footnote{https://swift.gsfc.nasa.gov/results/batgrbcat/}.
It contains 412 GRBs with the first burst GRB 050126 and the last one GRB 180510B (see Fig.~\ref{fig2}). The {\em Swift} data sample is preliminary. We are working on its completion.

The isotropic equivalent energies and luminosities $E_{iso}$ and $L_{iso}$ were calculated using the
luminosity distances (for measured redshifts $z$) in the Friedmann-Robertson-Walker model
assuming the Hubble constant $H_0=71\,\mathrm{km\,s^{-1} Mpc^{-1}}$, the matter density parameter
$\Omega_\mathrm{M}=0.27$, and the dark energy density parameter $\Omega_\mathrm{\Lambda}=0.73$.

The durations $T_{90,rest}$ of GRBs in their rest frames were calculated using
the cosmological time dilation as well as the energy-stretching dependence
(\cite[Fenimore et al. 1995]{Fe95}, \cite[M\'esz\'aros \& M\'esz\'aros 1996]{Me96})
 as $T_{90,rest}=T_{90,obs}/(1+z)^{1-k}$,
where $T_{90,obs}$ is the observed duration and $k=0.4$.

The preliminary clustering analysis based on Gaussian mixture models and Bayesian Information Criterion (BIC) (\cite[Kass \& Raftery 1995]{Ka95},
\cite[Mukherjee et al. 1998]{Mukh98})
reveals only two groups (BIC is maximal for two components) in the $E_{iso}$ vs. $T_{90,rest}$ plane in the {\em Fermi} data.
The separation occurs into the short/low-energy and long/high-energy clusters with difference in
BIC between the two components and one component $\Delta \mathrm{BIC}=17.6$ which suggests a very strong evidence in
favour of two components. Similarly, analysis in $L_{iso}$ vs. $T_{90,rest}$ plane in the {\em Fermi} data reveals only
two groups separated into the low-luminosity and high-luminosity with $\Delta \mathrm{BIC}=33.4$
suggesting a very strong evidence in favour of two components over one component. 
More than two components are not favoured in terms of BIC.

For {\em Swift}/BAT the data reveal three groups (BIC is maximal for 
three components) in the $E_{iso}$ vs. $T_{90,rest}$ plane. The separation 
occurs into the short/low-energy, long/high-energy 
and long/low-energy clusters with difference in BIC between 
the three components and the two components $\Delta \mathrm{BIC}=4.7$ 
which suggests a positive evidence in favour of three components. For 
the $L_{iso}$ vs. $T_{90,rest}$ plane in the {\em Swift} data, three groups 
(BIC is maximal for three components) are revealed. The separation 
occurs into the low-luminosity, 
intermediate-duration/high-luminosity and 
long/high-luminosity clusters with difference in BIC between 
the three components and the two components $\Delta \mathrm{BIC}=8.0$. 
This suggests a strong evidence in favour of three components. These 
results are preliminary and the completion of the data sample as well as 
 the statistical analysis is ongoing.

\section{Conclusion}

There are known several statistical tests, theories, ideas, modelling, etc... about the third or even about more subgroups.
A brief - never complete - survey was provided here. Summarizing these works it can be said that the existence of
any astrophysically different phenomenon - beyond the two (short/hard and long/soft) types - is further in doubt.
Both the intermediate subgroup and the possible subgroup of the low-luminosity long GRBs are not proven yet unambiguously.
Eventual further subgroups are also in doubt.

\end{document}